\documentclass{iopart}
\usepackage{graphics,epsfig ,color,graphicx} % figures en eps

\usepackage{iopams}

\begin{document}
\date{\today}
\title[Student self-training with multiple choice tests]{Multiple choice homework as a cost-effective and efficient tool for student self-training}
%Data treatment techniques]{A simple experiment highlighting data
%treatment techniques: determination of the relative dielectric constant of a
%material}
\author{G.L. Lippi \\}
\address{
Universit\'e C\^ote d'Azur, Institut de Physique de Nice,
UMR 7010 CNRS\\
1361 Route des Lucioles, F-06560 Valbonne, France}
\ead{gian-luca.lippi@inphyni.cnrs.fr}

\begin{abstract}
A self-training scheme geared at inducing students to improve their skills through independent homework is presented.  The motivation is to identify an inexpensive, yet effective tool for raising the competence level of students in the Fundamental Sciences (in particular Physics).  Since globally existing financial restrictions do not allow for extensive supervised work, a scheme is devised where the additional personal training is rewarded through bonuses in the grade, while safeguarding against the danger of cheating.  % associated with bonuses without direct supervision.
%
%it is necessary to offer (grade) compensation for the additional work from the student -- without overburdening the instructors -- while avoiding the danger of cheating which looms over unsupervised work.
Overburdening the instructors is avoided through the use of computer-based grading of homework, while a carefully chosen bonus plan, weighted by the grades obtained in supervised tests, counters the effects of potential cheating.
%
%
%A balance among the different needs and constraints is struck in the proposed scheme by using a computer-based grading of homework and a carefully chosen bonus scheme weighted by the grades obtained in supervised tests.
%
%with the use of computer-based grading coupled to weighted bonus points offered to students on the basis of supervised tests.
\end{abstract}
%\pacs{INSERT PACS}
\submitto{\EJP}
\maketitle

\section{Introduction}

As French Universities are not allowed to preselect students\footnote{Professional curricula, such as medicine, engineering, etc. are exempt from this and accept only a limited number of students each year on the basis of merit.} through entrance exams or through the assessment of previous performance (e.g., high-school results) the average level of classes is often rather poor.  The prohibition of any form of selection becomes an additional reason for the better students to run away from fundamental sciences in favour of applied ones (engineering in this case) where admission is based on merit and where the teaching resources and personalized attention are more generous.

A paradoxical situation arises, where the students who would need more follow-up -- due to their poorer initial background competence -- are relegated to curricula which do not receive sufficient resources to meet those needs.  Students spend a nonnegligible time in problem sessions, which should be expected to raise their competence level, but as the class size is at least 30 -- and often approaches 40 --, any form of the direly needed personalized help is impossible.   Nearly no provisions are made for closer follow-up, for instance by hiring Teaching Assistants, while the permanent personnel is assigned to large student groups.

The aim of this paper is to discuss a training technique devised to palliate this problem through a scheme which has proven effective with Physics students in the underprivileged environment outlined above.
The following discussion is therefore relevant to addressing the issue of how to raise the level of students who -- in average -- start their university studies with an insufficient background (in particular in mathematics) with the foreseeable consequences on the acquisition of the material as the curriculum progresses.  As such, the ideas I present will be most interesting to colleagues teaching in institutions where either no selection on merit is performed or which receive their students from the less favoured milieus.

One of the central points of the discussion is to seek a way of motivating the students and of providing easily accessible material for individual, independent (at home) training.  Due to lack of personnel/hour which can be dedicated to following the students' work, the aim is to find a way of efficiently improving the performance while keeping the time invested by the teaching personnel reasonably low.  Indeed, this time is not explicitely accounted for in the instructor's duties and comes from personal free time or from research time.

It is important to remark that due to this overall setting, a nonnegligible fraction of the enrolled students fails courses (at least at the first try), thus it is common to see class averages which are well below average.  While the consequent re-enrolment for the same course amounts to a waste of time (and a reduction in motivation for some students), the University fees are only nominal and do not place an additional financial burden on the student\footnote{The costs of tuition amount to approximately a couple hundred euros per year -- including some benefits such as access to sport facilities, reductions in transportation costs, health, etc.  For this reason Universities have to base their offer on funding provided almost exclusively by the government and cannot afford to offer teaching more adapted to the students' needs.  Discussions about raising the tuitions in special cases are under way and encounter strong resistance.  However, this is a political issue that does not enter into the topic of this paper, aimed at finding means of improving the current situation.}.  Nonetheless, the situation is highly unsatisfactory and, in spite of new measures proposed every year, the problem persists.  The training through the use of multiple choice tests appears to bring a small positive contribution to improving  students' competence, as explained in the following.

As a final remark, no grading on a curve is performed in the course I am referring to in this paper (and in general in no course in French Universities) and grades are given on the basis of a required competence level, rather than being adjusted on class average.  The problems outlined above translate into class averages which are consistently well below the passing mark (50\%).  In order to render the discussion more easily transferrable to the context of other systems (and countries), grades in this paper are normalized to their maximum and will be referred to either as percentages or numbers between $0$ and $1$\footnote{University-level French grades are given over 20 points, where 10 represents the passing grade.}.

The Student Self-Training (SST) scheme discussed below aims at improving the technical skills necessary for the quantitative analysis of the concepts acquired in the course of the Physics curriculum.  It does not represent a pedagogical technique for effectively presenting and acquiring concepts and as such is not in competition with the techniques (e.g., inverted classroom) that are being explored and tested for that purpose.  For the moment SST has been developed and used for the improvement of students' mathematical skills, but it is envisageable to extend it to complement the standard, weekly class homework directly pertaining to the topic of the course.

\section{Self-training scheme}\label{scheme}

The idea of the SST scheme is to motivate students to perform additional work at home to improve their mathematical skills, which are normally the first and foremost -- though not unique -- stumbling block in their path towards the successful acquisition of expertise in Physics.
\footnote{Additional support material is provided to the students in the form of ``reminder'' of techniques and fundamentals which are considered as prerequisite -- and which should have been acquired prior to enrolling in the course.  These materials include, in addition to specific writeups, web resources which the student can (and hopefully will) consult independently.}.  This homework, geared at improving mathematical skills through exercises specifically matching the kinds of calculations appearing in course--related problems, is added to the required one, to laboratory work, to exam preparation, etc.  Thus, it is necessary to provide incentives which may induce the students to accept the challenge.  The unsupervised nature of the work performed at home limits the way credit can be assigned for this additional effort because of the potential for cheating, which has to be dealt with in a creative way.

The homework consists of a set of problems (same for all students) placed on a web page freely available.  The interested students are required to download the text, print the answer matrix, fill-in the correct answers and return the papers to the instructor before the mid-term exam (at the latest just before the beginning of the test).

The conditions under which the SST is conducted are the following:
\begin{itemize}
\item[$\bullet$] The homework is unsupervised and the student is allowed to consult books, references and web resources in order to learn;
\item[$\bullet$] Discussion with fellow students are acceptable, but simply copying the result from someone else (or from an available solution) is not;
\item[$\bullet$] No enforcement of rules is reasonably possible, thus it is impossible to detect cheating, if present.
\end{itemize}

\begin{figure}[ht!]
\includegraphics[width=0.9\linewidth,clip=true]{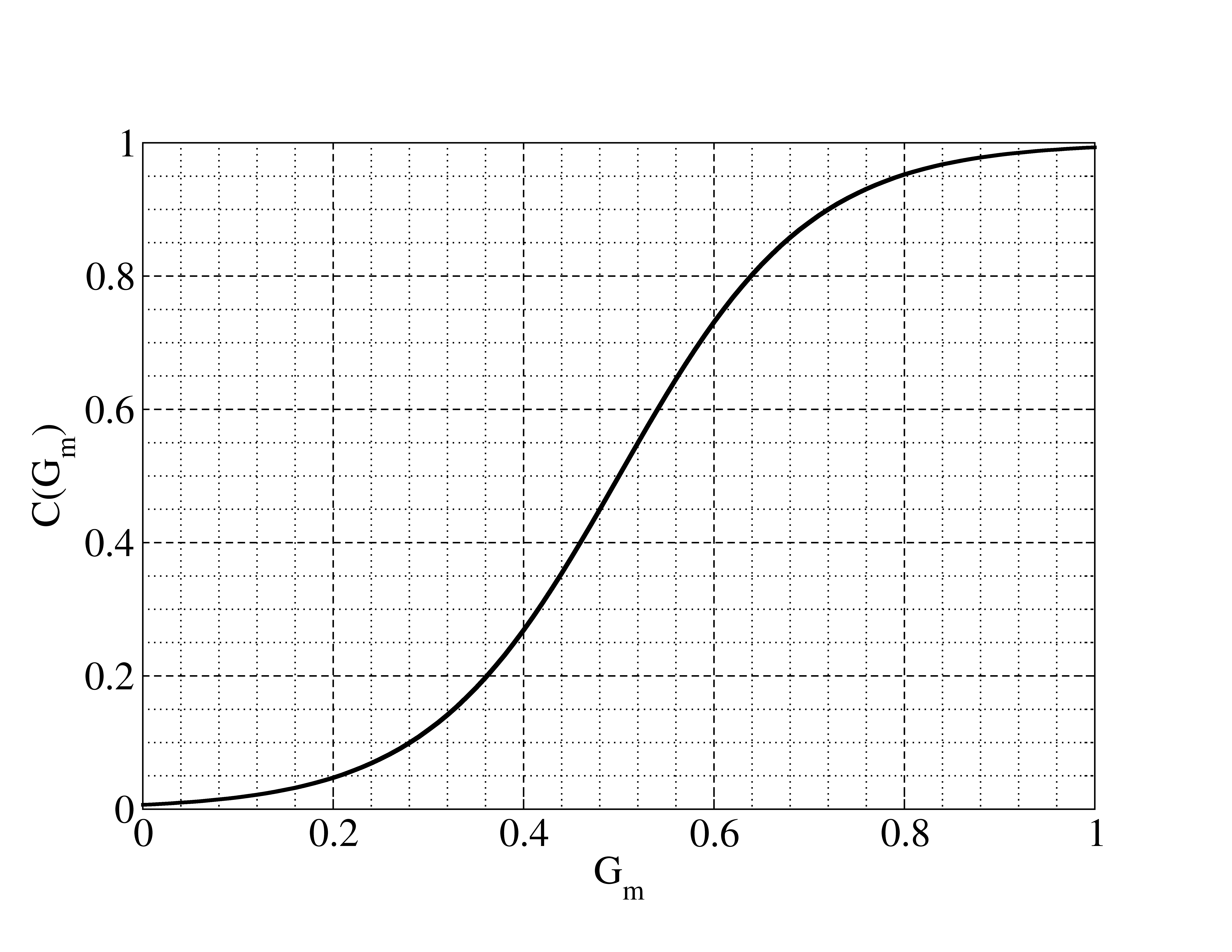}
\caption{
Weight coefficient $C(G_m$) for cumulating the benefit of the homework grade, $G_h$, onto the mid-term grade $G_m$.  The full credit ($G_{h,max}$) assigned to $G_h$ is a free parameter which can be adjusted as a function of the quantity of exercises given, their difficulty, the amount of time allotted and the maximum benefit to be given to the students.  In the examples which follow $G_{h,max} = 0.25$.  The function $C(G_m)$ is plotted, and used, with $\alpha = 5$.
} 
\label{sigmoid} 
\end{figure}

The last point represents the potentially weak link in the scheme.  However, cheating is rendered useless by the way the incentives for students are offered:
\begin{itemize}
\item[$\bullet$] While the homework is voluntary, in addition to the obvious advantage coming from additional training, students derive a direct benefit from their efforts from the attribution of ({\it conditional}) bonus-points, as described below;
\item[$\bullet$] The points acquired in the homework are added to the mid-term grade according to a weighting function $C(G_m)$ (sigmoid):
\begin{eqnarray}
\label{corrgrade}
F_m = G_m + C(G_m) \times G_h,
\end{eqnarray}
where $F_m$ stands for the {\it final} mid-term grade (comprehensive of the bonus), $G_m$ for the grade obtained in the written mid-term test, $C(G_m)$ is the weight and $G_h$ is the grade obtained in the homework.\\
The {\it sigmoidal} function $C(G_m)$ used in this scheme is
\begin{eqnarray}
C(G_m) & = & \frac{1 + \tanh (\alpha G_m)}{2}
\end{eqnarray}
where $\alpha$ is a coefficient controlling the slope of the hyperbolic tangent\footnote{Depending on the choice of value for $\alpha$ -- cf. caption of Fig.~\ref{sigmoid} -- the accessible range of weights does not cover the entire interval $[0,1]$.  However, as clear from Fig.~\ref{sigmoid}, the values for $C(G_m = 0)$ and $C(G_m = 1)$ are close enough to 0 and 1, respectively, to render the error negligibly small.} (cf. Fig.~\ref{sigmoid});
\item[$\bullet$] As clear from the weight coefficient $C$, students who gain 50\% of the points in the mid-term ($G_m = 0.5$) cumulate 50\% of the homework points ($C(0.5) = 0.5$).  The steep growth of the sigmoid between $G_m = 0.25$ (where practically no homework credit is accumulated) and $G_m = 0.75$ (where most of the homework credit is cumulated) encourages students to strongly improve their performance.  The objective of the exercises is exactly to achieve this goal.
\item[$\bullet$] Students who may underperform in the mid-term, in spite of homework training,  will anyway benefit from the practice on the longer term since it will help them in the final exam (as witnessed in the real setting).
\item[$\bullet$] Students who may have cheated in the homework by copying the answers do not draw any benefits from the grade $G_h$ they have obtained, since it is extremely unlikely that they will be able to perform well in the exam (leading to $C \approx 0$).  Thus their ``effort'' will have been wasted.
\end{itemize}

This summary clearly illustrates the benefits of the grading scheme and its resilience against cheating.

\section{Homework preparation and time investment}

The homework set is constructed on the basis of a program (Auto-Multiple-Choice, AMC) designed for tests based on multiple choice questions and briefly described below.  The reason for this choice is the automatization of the grading of this additional homework which transforms an impossibly lengthy task for instructors, with the constraints outlined above (lack of personnel), into a reasonable one.  Given that, as previously discussed, the homework set can be reused from one year to another, the development of the set of questions becomes an investment worth the required time.

AMC is not the only suitable program, but it is the one used for these tests and the one for which some useful information can be given here.  The program offers, among many other options which are not relevant for this discussion, automatic grading of the assignment from a scanned image of the matrix of answers returned by the student, with the boxes corresponding to the correct results blackened out in pen ink.  

The AMC programme is based on a graphical interface which uses LaTeX as underlying language, but with specific commands and functions specifically developed for the purpose.  No knowledge of LaTeX is required, as AMC is provided also with its own set of commands (the user chooses between those or LaTeX commands).  AMC is not specifically developed for the sciences, thus it is quite versatile and offers a large palette of options.  As such, it is not always easy to use and the fairly extensive manual requires, on occasion, some effort to interpret the meaning of specific functionalities.  AMC is free and can be installed on Windows-based systems, as well as on most Linux releases.  For more information, the potentially interested user should look at the AMC web page~\cite{AMC}.  

The AMC option chosen to prepare the homework is the one where multiple correct answers are allowed, which automatically implies the possible absence of any correct answer among the proposed set.  Other existing options, such as answer scrambling or randomization, are not used here.

In the implementation I have set up for self-training each question proposes a certain number of answers, with a minimum of at least 6 (more if possible) to render guessing too unlikely to be useful.  An important, and time-consuming, part of the work is preparing incorrect answers in sufficiently large numbers for each question.  Typically, I offer at least two groups of answers:  one containing the correct answer(s), with the addition of a few variations consisting of small changes (a coefficient, for instance) for the purpose of testing the student's ability to perform the full calculation correctly, down to the final detail; at least a second group of answers (more if possible) based on common mistakes (even gross ones) with, again, some variations.   Occasionally, one question is proposed without any correct answer to keep the student from assuming that there must be a right one and therefore trying to ``bend'' the calculation to try and find it.  This choice is made to reinforce the student's confidence in personal work, rather than in external suggestions.

The AMC option of attributing negative points to incorrect answers is used to discourage the blackening of random boxes in the hope of hitting on the correct answer.  Since the number of points (both positive and negative) is attributed to each individual answer by the instructor at the time of generating the code, it is possible to sanction entirely wrong answers more strongly than answers which are a slight variation of the correct one (e.g., due to an incorrect coefficient but correct functional dependence).

A very important point, in light of a possible interest in using the technique in other settings, is an estimate of the time to be invested into the project to set it up the first time and to run it.  This estimate is not an easy task, as the details depend crucially on the available facilities (scanner speed and quality, connection, computer speed and memory, etc.), but is offered in the hope of providing at least an idea of the required effort and help potentially interested colleagues in making their choice.

\begin{table}
\caption{
Number of students, $N_e$, enrolled in the Waves class (3$^{\rm rd}$ semester) in the last four academic years.  For the 2017-18 class, $N_1$ is the number of students handing in the homework, while $N_2 = N_e - N_1$ stands for the students who have chosen not to do so.
}
\label{studNo}
\lineup
\begin{indented}
\item[]\begin{tabular}{c c c c}\br 
Year & $N_e$ & $N_1$ & $N_2$  \\  \mr 
2014-15 & 69 & -- & --  \\ 
2015-16 & 63 & -- & --  \\
2016-17 & 71 & -- & --  \\
2017-18 & 68 & 21 & 47 \\ 
\br
\end{tabular}
\end{indented}
\end{table}

First, one needs to invest into learning how to use AMC.  A good mastery of the program can be acquired by someone without previous experience in about a week; less for users familiar with LaTeX and intending to just learn the basics for this project.  However, this is a one-off investment which can also be rendered profitable through its use in other settings (ex. preparation of true in-class tests, even in randomized form).

The preparation of questions and of answers (mostly wrong ones) is something that only each instructor can evaluate, since it depends on personal factors and on class topics.  This time would be nearly the same if the homework were prepared for standard (i.e., by hand) correction -- up to the preparation of the wrong answers.  

The typesetting, including debugging which again depends on the preparer's skills with the program, may take approximately one day for 30 questions.  This part is also a one-time investment, since one can later reuse the same homework for the years to come.  Placing the questions on a stable web site ensures the durability of the posting while the time required for this operation can be considered negligible, compared to the rest.

The recurrent time investment comes from collecting the homework, scanning the sheets, grading and returning the results to students (preferably by email, since the correction is electronic).  From experience, one can expect to scan, grade and return the files by email in approximately 30 minutes for every 15 students without using the mail server which AMC offers (useful mostly when large number of students are involved, as in introductory courses).  This is an estimate based on personal experience and depends on the actual material (scanner, computer, etc.), but it conveys the order of magnitude of the required time.  Compared to the benefits, the investment is profitable and sustainable.

\section{Analysis}

The SST scheme was implemented with the class 2017-18 of the Waves course, held by the author for 3$^{\rm rd}$ semester Physics students at the Universit\'e de C\^ote d'Azur.  Of the 68 students enrolled in the class, only 21 chose to return the homework.  

In Table~\ref{studNo}, the global class results are compared, to test for variations in the average grade, to those of three previous years (from 2014-15 to 2016-17) when no form of additional optional homework was offered.  Details on number of students are given in Table~\ref{studNo}. 

\begin{figure}[ht!]
\includegraphics[width=0.9\linewidth,clip=true]{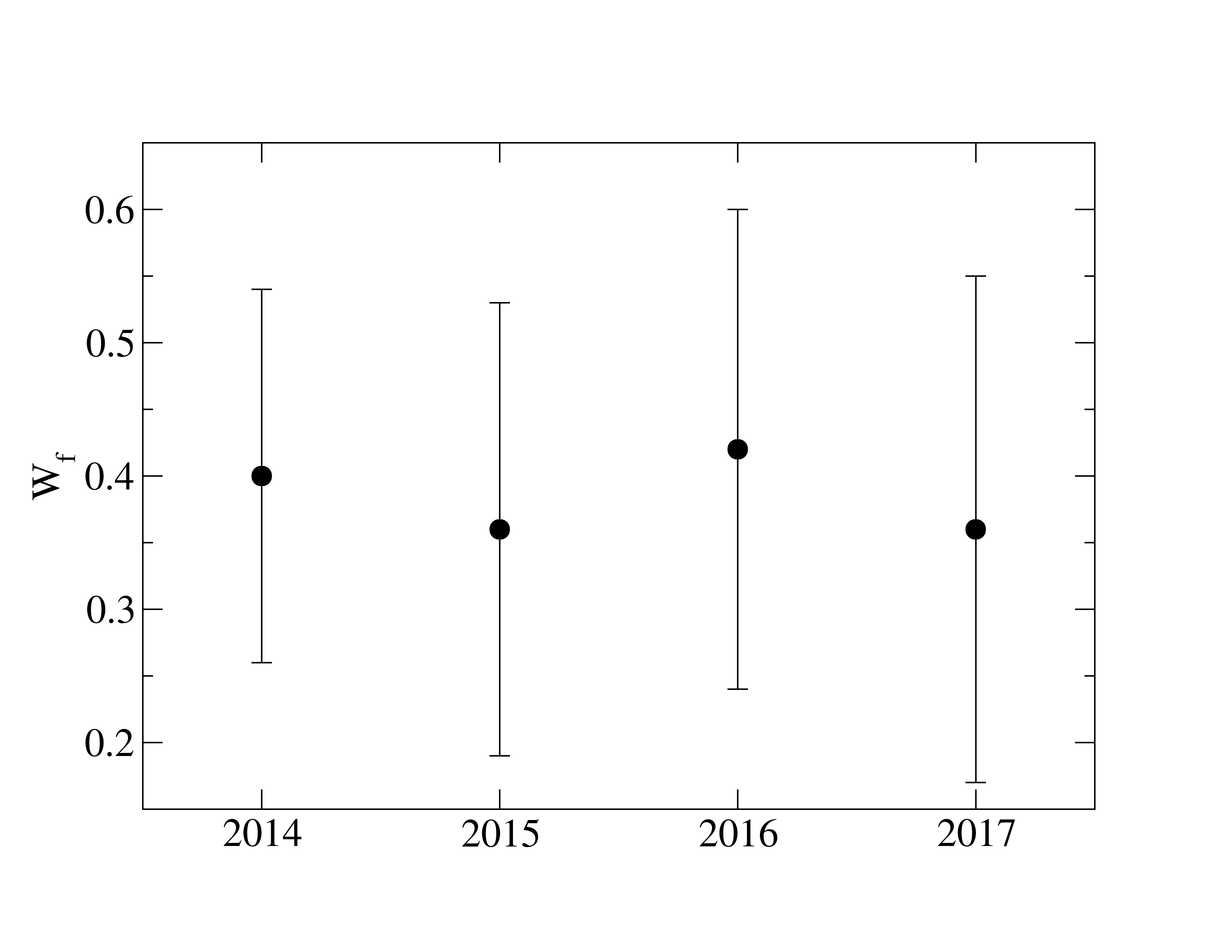}
\caption{
Final grade $W_f$ with standard deviation for the students enrolled in the course for the academic years indicated in the abscissa (only the enrollment year is marked, for simplicity).  Notice that the average grade fluctuations from one cohort to the next are rather small and that, overall, the last group performed comparably to (even somewhat less well than) the previous years.
} 
\label{nfinal} 
\end{figure}

Fig.~\ref{nfinal} graphically shows similar information:  the class average for the different years, accompanied by its standard deviation.  The results show that the variations from one year to the next are rather small, thus we can exclude a bias coming from a better-prepared cohort of students, or from a different degree of difficulty in tests.  Indeed, the overall performance of the 2017-18 class was even somewhat poorer.

\begin{figure}[ht!]
\includegraphics[width=0.9\linewidth,clip=true]{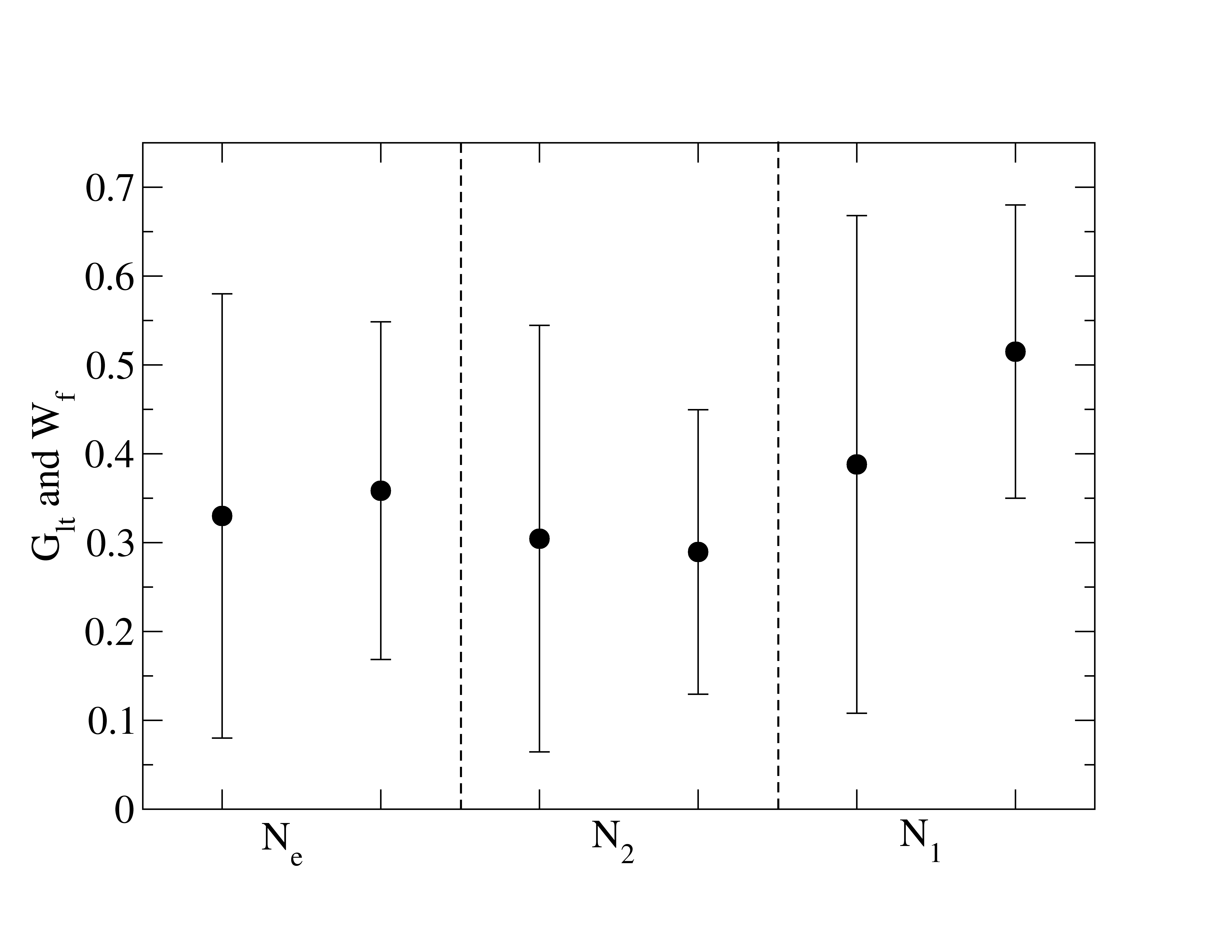}
\caption{
Level test grade ($G_{lt}$, left) and final grade ($W_f$, right) displayed with its standard deviation for:  $N_e$ -- all students; $N_2$ -- students not having returned the homework; $N_1$ student who returned the homework.  The dashed vertical lines separate the grades by group, as indicated in the labels on the abscissa.
} 
\label{progress} 
\end{figure}

The third column in Table~\ref{prog} gives the normalized average grade for all students (first row), for the students who returned the homework (group 1, second row) and for those who did not do the homework (group 2, third row).  The same data is graphically reported, with error bars, in Fig.~\ref{progress}.  

In 2017-18, in addition to the SST option, a 30-minutes-long multiple-choice test  was administered to probe the mathematical competence on the needed basic concepts; its scope was to allow for a quantification of the students' progress.  

Students who later chose to do the homework scored sensibly higher (Fig.~\ref{progress}  -- values from Table~\ref{prog}) in this initial test ($G_{lt} = 0.39$) than their remaining colleagues ($G_{lt} = 0.31$):  this may be an indication of the willingness of this group to engage in the additional training.  The final grade shows, however, a deepening of the hiatus between the two groups:  the first one gained $0.13$ points, while the other one lost $0.02$ points between the level test and the final grade (cf. last column in Table~\ref{prog}).

\begin{table}
\caption{
Number of students, $N_e$, enrolled in the Waves class (3$^{\rm rd}$ semester).  For the 2017-18 class, the first line contains the information for all students, the second for those of group 1 (i.e., those having handed in their homework), while the last one corresponds to group 2 (no homework).  The second column reports the normalized grade for the level test administered at the beginning of the course, the third the final normalized grade, and the last the progression.
}
\label{prog}
\lineup
\begin{indented}
\item[]\begin{tabular}{c c c c}\br 
Group & $G_{lt}$ & $W_f$ & $W_f - G_{lt}$  \\  \mr 
All & 0.33 & 0.36 & +0.03  \\ 
1 & 0.39 & 0.52 & +0.13  \\
2 & 0.31 & 0.29 & -0.02  \\
\br
\end{tabular}
\end{indented}
\end{table}

It is important to remark that the average contribution of the homework grade directly onto the final grade (i.e., the average of term $C(G_m) \times G_h$ in equation~\ref{corrgrade}) is $0.03$ (only for group 1).  This is due to the additivity of the offered bonus (cf. discussion in section~\ref{scheme}).  Thus, the remaining  average gain ($0.1$) can be reasonably interpreted as representing the average student progression coming from the additional training.  Since this amounts to 10\% of the maximum grade, the progress is substantial.

%While the bias coming from the better initial level of the students belonging to group 1 may explain the choice of additional training, the 10\% improvement gained over the semester is a nonnegligible influence which merits attention.

\section{Conclusions}

While in an ideal setting one would have students arriving at University with a sufficiently strong background and skills to independently fill the (hopefully small) gaps which may possibly remain in their previous education, the current situation proves the contrary.  Even in the presence of a web environment which provides a broad self-training offer, it has become necessary to convince and guide students to improve their technical knowledge, in order to complete a degree in science.  The current global economic situation renders this  potentially very expensive action unfeasible, thus alternative schemes for self-training must be sought.

The challenge is to motivate students in doing what has always been known to be the necessary step for progress (exercise) without incurring into potential cheating problems or into very high costs, which cannot be met by the Institutions.
The SST scheme proposed in this paper, based on computer-graded multiple choice answers, fulfills the needs of both sides:  a low-cost investment for instructors (at zero-cost for the Universities) and an effective, cheat-proof progression tool designed to encourage students to engage in the process.  

It would be very interesting to extend this experiment to different Institutions belonging to different countries and including various specialties.  Colleagues interested in such ideas are welcome to contact the author for discussions and for potential future collaborations.

\ack
Technical computer help from G.P. Puccioni is gratefully acknowledged.  The author is  grateful to the developers of AMC and to the whole team which supports its implementation on different platforms.  This work benefitted from the AMC version installed in a Linux Fedora operating system.

\end{document}